\documentclass[fleqn,usenatbib]{mnras}

\usepackage{newtxtext,newtxmath}

\usepackage[T1]{fontenc}
\DeclareRobustCommand{\VAN}[3]{#2}
\let\VANthebibliography\thebibliography
\def\thebibliography{\DeclareRobustCommand{\VAN}[3]{##3}\VANthebibliography}

\usepackage{amsmath}	
\usepackage{natbib}
\usepackage{bm}
\usepackage{dcolumn}
\usepackage{enumerate}
\usepackage{multirow}
\usepackage{inputenc}
\usepackage[caption=false]{subfig} 
\usepackage{fontenc}
\usepackage{float}
\usepackage{graphicx, rotating, threeparttable}
\usepackage{xcolor}
\usepackage{geometry}
\usepackage{array, multirow}
\usepackage[export]{adjustbox}
\usepackage{subfig}

\hypersetup{
  colorlinks = true,
  linkcolor  = black
}

\title[]{Singling out modified gravity parameters and datasets reveals a dichotomy between Planck and lensing}

\author[]{
Cristhian Garcia-Quintero,$^{1}$\thanks{E-mail: gqcristhian@utdallas.edu}
and Mustapha Ishak$^{1}$\thanks{E-mail: mishak@utdallas.edu}
\\
$^{1}$Department of Physics, The University of Texas at Dallas, Richardson, Texas 75080, USA
}

\date{Published June 2021}

\pubyear{2020}

\begin{document}
\label{firstpage}
\pagerange{\pageref{firstpage}--\pageref{lastpage}}
\maketitle

\begin{abstract}
An important route to testing General Relativity (GR) at cosmological scales is usually done by constraining modified gravity (MG) parameters added to the Einstein perturbed equations. Most studies have analyzed so far constraints on pairs of MG parameters, but here, we explore constraints on one parameter at a time while fixing the other at its GR value. This allows us to analyze various models while benefiting from a stronger constraining power from the data. We also explore which specific datasets are in tension with GR. We find that models with ($\mu=1$, $\eta$) and ($\mu$, $\eta=1$) exhibit a 3.9-$\sigma$ and 3.8-$\sigma$ departure from GR when using Planck18+SNe+BAO, while ($\mu$, $\eta$) shows a tension of 3.4-$\sigma$. We find no tension with GR for models with the MG parameter $\Sigma$ fixed to its GR value. Using a Bayesian model selection analysis, we find that some one-parameter MG models are moderately favored over $\Lambda$CDM when using all dataset combinations except Planck CMB Lensing and DES data. Namely, Planck18 shows a moderate tension with GR that only increases when adding any combination of RSD, SNe, or BAO. However, adding lensing diminishes or removes these tensions, which can be attributed to the ability of lensing in constraining the MG parameter $\Sigma$. The two overall groups of datasets are found to have a dichotomy when performing consistency tests with GR, which may be due to systematic effects, lack of constraining power, or modelling. These findings warrant further investigation using more precise data from ongoing and future surveys.

\end{abstract}




\section{Introduction}
%
General Relativity has survived over a hundred years in the same original formulation that Einstein had proposed in 1915 \cite{Einstein1915}. 
It has been the subject of intensive tests in the solar system and the strong regime, see, e.g., \cite{Will1993,Will2018,Berti:2015itd}. In the last two decades, GR has become the subject of tests and studies in cosmology that have become possible due to the rapidly growing amount of cosmological data available from ongoing and planned surveys,  as we discuss below. 

Cosmological tests of gravity theory is a justified endeavor in its own right but it is often invoked in relation to the problem of cosmic acceleration and the dark energy associated with it, see, e.g., the following reviews for various formulations and perspectives on the problem  \cite{Weinberg1988CC,Padmanabhan2003,Sahni2000,Carroll2001CC,Peebles2003,Copeland2006,Ishak:2005xp}. Interestingly, gravity theory affects not only the rate of expansion of the universe but also the rate at which structures, like clusters and superclusters, form and grow in the cosmic sub-stratum. It turns out that this growth rate of structure is sensitive to the underlying theory of gravity and thus is useful in designing tests of gravity at cosmological scales. So even if two theories can have the same degenerate cosmic expansion history, the growth rate of structure can still serve to discriminate between the two theories, see for example some early papers \cite{Linder2005,2006-Ishak-splitting}.
Accordingly, a promising route to test deviations from GR at cosmological scales is to model such deviations at the level of the perturbed Einstein's equations, see, e.g. the reviews \cite{Clifton:2011jh,Koyama:2015vza,Joyce:2016vqv,Ishak:2018his} and references therein. One popular approach is to add  phenomenological parameters to the perturbed equations that represent some physical effect such as an enhanced or suppressed growth rate of structure.  
This has been done for spatially flat and curved models \cite{GongCurved2009,DossettMGC2012}.
Often, two parameters are introduced with one of them representing matter clustering or gravitational coupling and the other parameter is related to light propagation in the perturbed spacetime, see, e.g. \cite{DossettFOM2011,2018MG-Gannouji,2019MG-Kazantzidis}. Depending on the specific implementation, such parameters will take the value of zero or unity in the case of GR, but will deviate from those values in the case of modified gravity (MG). The data is then used in order to attempt to accurately constrain those parameters and test whether they are consistent or not with their GR values, see, e.g. \cite{JoudakiEtAl2017,Nesseris:2017vor,Blake:2020mzy}. Other approaches to testing GR at cosmological scales include looking at specific MG models, see e.g. \cite{Burrage2018,DGP}, looking into the velocity fields, see e.g. \cite{Hellwing2014,Carlesi2017} or considering beyond scalar perturbations, see e.g. \cite{Lin2016,Bernal2020}

While previous papers have focused on constraining two parameters at the same time, we focus in this paper on constraining only one single parameter at a time using the data available. We also investigate what datasets or combinations of datasets are consistent (or not) with GR by using specific  combinations of datasets. Among the motivations for this is that we want to use all the available constraining power against one parameter at a time to try to obtain more stringent constraints. A second motivation is to explore what various types of datasets have to say about each individual parameter and the corresponding deviation from GR.        

The paper is designed as follows. Section 1 is an introduction. In section 2, we provide the underlying perturbed equations and corresponding parameterizations. The datasets used are described in section 3, while the results and their analysis are presented in section 4. We conclude in section 5. An appendix briefly describes a Python wrapper for the system \texttt{ISiTGR} that is used to test GR at cosmological scales and that is released with this paper.

\section{MG equations and parameterization\label{sec:MG-equations}}
A convenient approach in order to test deviations from GR at cosmological scales is to modify the equations that affect the gravitational potentials at the level of linear perturbations. The starting point is to consider the scalar perturbed flat Friedmann-Lema\^{i}tre-Robertson-Walker (FLRW) metric
\begin{equation}
ds^2=a(\tau)^2[-(1+2\Psi)d\tau^2+(1-2\Phi)\delta_{ij}dx^i dx^j].
\label{eq:line-element}
\end{equation}
It can be shown that a subsequent treatment through the perturbed Einstein equations leads to a relativistic Poisson equation
\begin{equation}
k^2\Phi = -4\pi G a^2\sum_i \rho_i \Delta_i,
\label{eq:Poisson}
\end{equation}
and provides a second equation that relates the two gravitational potentials given by 
\begin{equation}
k^2(\Psi-\Phi) = -12 \pi G a^2\sum_i \rho_i(1+w_i)\sigma_i,
\label{eq:2ndEin}
\end{equation}
represented in Fourier space. The formalism we are following is based on modifying these equations by adding phenomenological parameters that are sensitive to different cosmological probes and that in general can be functions of both scale and time. The MG parameter related to the behaviour of non-relativistic particles and which modifies the growth of structure is defined by
\begin{equation}
k^2\Psi = -4\pi G a^2\mu(a,k) \sum_i\left[\rho_i\Delta_i+3\rho_i(1+w_i)\sigma_i\right],
\label{eq:definition_mu}
\end{equation}
obtained by use of (\ref{eq:Poisson}) and (\ref{eq:2ndEin}). Furthermore, it is expected that in the current universe $\Psi=\Phi$ for a negligible shear stress. Then, a significant deviation from $\Psi=\Phi$ should be an indication of an underlying theory different from GR. Hence, we can define a parameter $\eta(a,k)$ that tests the equality between the two potentials by replacing (\ref{eq:2ndEin}) with
\begin{equation}
k^2[\Psi-\eta (a,k)\Phi] = -12 \pi G a^2\mu (a,k)\sum_i \rho_i(1+w_i)\sigma_i.
\label{eq:definition_eta}
\end{equation}
Now, whereas $\mu(a,k)$ quantifies the strength of the gravitational interaction, it is convenient to define another MG parameter which is sensitive to the deflection of light in a given matter field. Such a parameter is defined from (\ref{eq:Poisson}) and (\ref{eq:2ndEin}) as
\begin{equation}
k^2(\Phi+\Psi)=-4\pi G a^2 \Sigma (a,k) \sum_i\left[2\rho_i\Delta_i + 3\rho_i(1+w_i)\sigma_i\right].
\label{eq:definition_Sigma}
\end{equation}
Thus, the MG parameter $\Sigma(a,k)$ is proportional to the Weyl potential $(\Phi+\Psi)/2$ and hence it quantifies deviations from GR associated with the response of massless particles in the gravitational lensing effect. 

Among the different MG parameterizations, the most commonly used in the last years have been based on combinations of MG parameters such as ($\mu$, $\eta$) and ($\mu$, $\Sigma$), used for example in \cite{Planck2015MG} and \cite{DESMG2018} respectively. Finally, it is worth noticing that we recover GR when $\mu=\eta=\Sigma=1$. 

\begin{table*}
\begin{tabular}{ m{2.5cm} | l }
\hline
Datasets & Description  \\ \hline
\multicolumn{2}{l}{1. Individual datasets}  \\ \hline
\quad \multirow{2}{*}{TTTEEE} & Planck high-$\ell$ temperature and polarization spectra, and low-$\ell$ temperature Commander likelihood \\
& \cite{Planck2018} \\
\quad lowE & Planck low-$\ell$ SimAll E-mode polarization spectra likelihood \cite{Planck2018} \\
\quad CMBL & Light deflection measurements from the CMB \cite{Planck-2018-lensing} \\
\quad SNe & Pantheon supernovae type Ia compilation \cite{Pan-STARRS-2017} \\ 
\quad 6dFGS & BAO measurements from the 6dFGS \cite{BeutlerEtAl2011} \\
\quad MGS &  BAO measurements from the SDSS MGS  \cite{RossEtAl2014} \\
\quad BOSS &  BAO consensus results from BOSS DR12 \cite{AlamEtAl2016} \\ 
\quad Ly-$\alpha$ & BAO measurements from the correlation of Lyman-$\alpha$ forest absorption and quasars \cite{2019-BAO-lyalpha-quasar-Blomqvist} \\
\quad RSD  & SDSS III galaxy clustering data from BAO spectroscopic survey \cite{AlamEtAl2016} \\
\quad BBN & One percent determination of the primordial deuterium abundance \cite{Cooke2018}  \\
\quad HST & Hubble Space Telescope local measurements of $H_0$ \cite{Riess2019H0} \\
\quad DES  & Dark Energy Survey Year 1 clustering and lensing analysis \cite{DES2017} \\ \hline
\multicolumn{2}{l}{2. Combined datasets} \\ \hline
\quad P18 & TTTEEE+lowE \\
\quad BAO & BOSS+6DFGS+MGS+Ly-$\alpha$ \\
\quad SBB & SNe+BAO+BBN \\
\hline
\end{tabular}
\caption{Summary of the datasets used in this work. We list the datasets individually and combinations of them that are relevant in our analysis.}
\label{Table:DataSets}
\end{table*}
\begin{table}
\begin{center}
\scriptsize 
 \begin{tabular} {c|c|c|c}
\hline
 MG parameterization & Free parameters & Relationships & Discussed \\
(see equations (\ref{eq:explicit-form_mueta1}-\ref{eq:explicit-form_muSigma2}))  &  &                   & in e.g. Refs.\\
 \hline
($\mu$, $\eta$) & $E_{11}$, $E_{22}$ & $\Sigma = \mu (\eta+1) / 2$ &  \cite{Planck2015MG} \\ 
($\mu$, $\Sigma$) & $\mu_{0}$, $\Sigma_{0}$ & $\eta=(2\Sigma - \mu)/\mu $ & \cite{SimpsonEtAl2013} \\  
($\mu$, $\eta=1$) & $E_{11}$, $E_{22}=0$ & $\Sigma=\mu$ & \cite{LInder2017OmegaDE2} \\ 
($\mu=1$, $\eta$) & $E_{11}=0$, $E_{22}$ & $\Sigma = (\eta + 1)/2$ & \cite{Linder2020-limitedMG} \\ 
($\mu=1$, $\Sigma$) & $\mu_{0}=0$, $\Sigma_{0}$ & $\eta = 2\Sigma-1$ & \cite{Linder2020-limitedMG} \\  
($\mu$, $\Sigma=1$) & $\mu_{0}$, $\Sigma_{0}=0$ &$\eta = (2 - \mu)/\mu$ & \cite{Linder2020-limitedMG} \\  
\hline
\end{tabular}
\end{center}
\caption{MG models considered in our analysis. We derive  constraints on models with two MG parameters. We also derive the constraints on these models when setting one of the two MG parameters to its GR value and observe the changes in the constraints from different probes.}
\label{Table:MG-models}
\end{table}

 There have been several attempts to find an adequate functional form for the MG parameters that can efficiently quantify the signatures of departure from GR. In this work, we follow what has been done by recent collaborations by adopting a time-dependent functional form for the MG parameters that is proportional to the dark energy density parameter $\Omega_{\text{DE}}(a)$. 

In order to test these extended models based on the addition of MG parameters to the $\Lambda$CDM model, we use the Integrated Software in Testing General Relativity (\texttt{ISiTGR}) \cite{ISITGR,ISiTGR-CGQ}. \texttt{ISiTGR} is a patch to \texttt{CosmoMC} \cite{COSMOMC} and \texttt{CAMB} \cite{CAMB} which is able to calculate predictions and do MCMC sampling for MG models based on current data in cosmology. As coded in \texttt{ISiTGR} and following the original conventions, we model the MG parameters as 
\begin{equation}
\mu(a)=1+E_{11}\Omega_{\text{DE}}(a)
\label{eq:explicit-form_mueta1}
\end{equation}
and
\begin{equation}
\eta(a)=1+E_{22}\Omega_{\text{DE}}(a),
\label{eq:explicit-form_mueta2}
\end{equation}
for the ($\mu$, $\eta$) parameterization. If we define the values of the MG parameters at present time as $\mu_0=\mu(z=0)$ and $\eta_0=\eta(z=0)$, then $\mu_0$ and $\eta_0$ can be determined from $E_{11}$ and $E_{22}$ alongside the dark energy density parameter at today. Similarly for the ($\mu$, $\Sigma$) parameterization, we use
\begin{equation}
\mu(a)=1+\mu_{0}\frac{\Omega_{\text{DE}}(a)}{\Omega_\Lambda}
\label{eq:explicit-form_muSigma1}
\end{equation}
and
\begin{equation}
\Sigma(a)=1+\Sigma_{0}\frac{\Omega_{\text{DE}}(a)}{\Omega_\Lambda},
\label{eq:explicit-form_muSigma2}
\end{equation}
where the only difference with respect to the ($\mu$, $\eta$) parameterization is the normalization factor using the current dark energy density. Similarly, here $\Sigma_0=\Sigma(z=0)$ is defined as the value of the MG parameter at present time. We opt not to use scale dependence in the MG parameters since this seems to lead to weak constraints. Also, it is important to mention that we recover GR when $E_{11}=E_{22}=0$ or $\mu_0=\Sigma_0=0$. Some differences between $\Lambda$CDM and these MG models are shown in Appendix \ref{appendix:ISiTGR}, where we plot the Cosmic Microwave Background (CMB) angular power spectrum and matter power spectrum for some combinations of the MG parameters by using some features of the new Python wrapper for \texttt{ISiTGR}. The proportionality of the MG parameters to $\Omega_{\text{DE}}$ is widely used in MG studies and literature but it was pointed out in \cite{Linder2017} that it can incorrectly estimate the observables and such a parameterization can have limitations in capturing deviations from GR. Our goal here, though, is to examine how different combinations of data constrain each of the MG parameters within such widely-used parameterizations. We will follow up with a separate full study using MG binning methods that have the potential to address some of the concerns of \cite{Linder2017} which are beyond the scope of this paper.  
\begin{figure*}
\begin{tabular}{c c}
{\includegraphics[width=8.1cm]{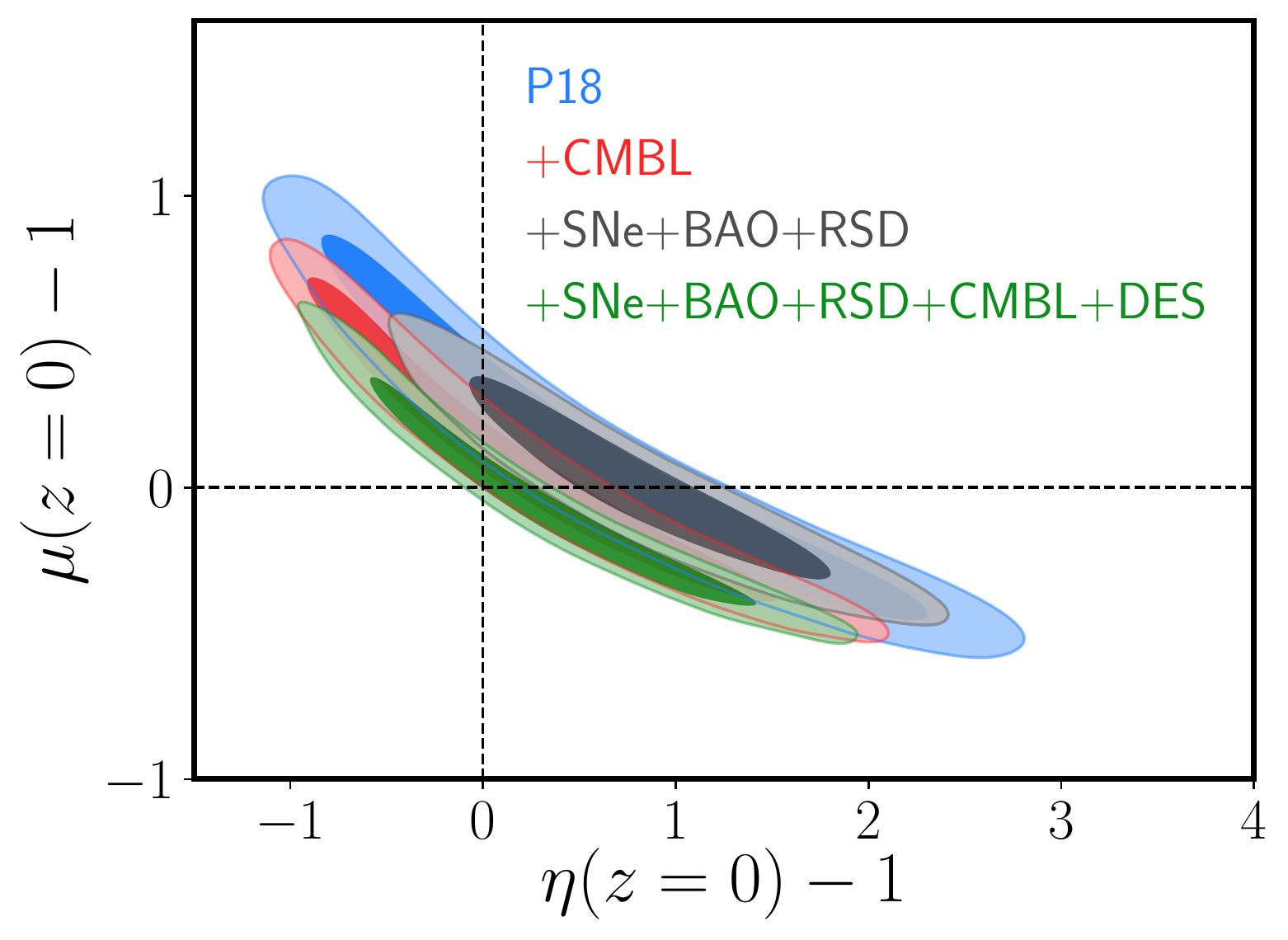}} &
{\includegraphics[width=8.1cm]{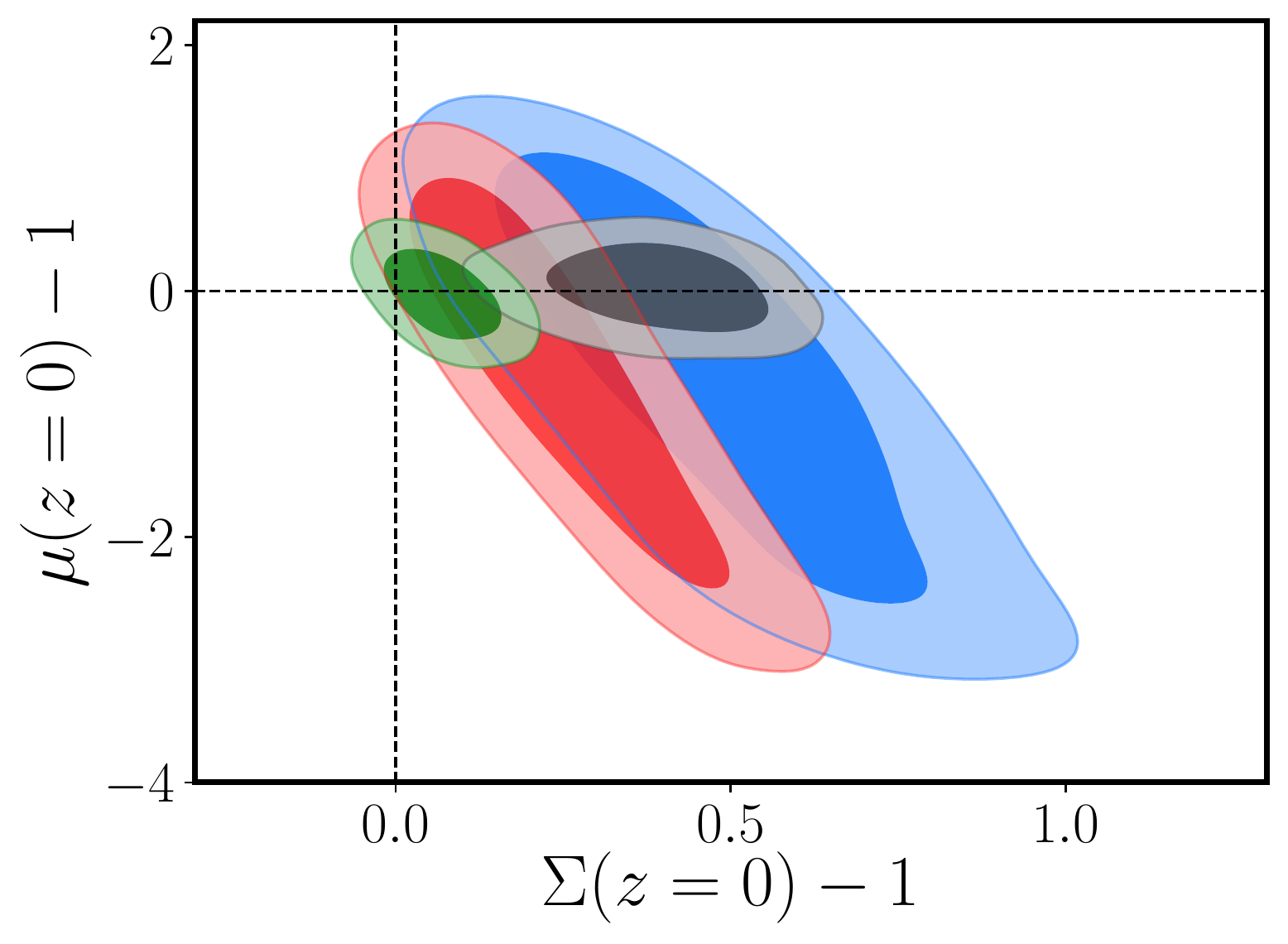}} 
\end{tabular}
\caption{68\% and 95\% confidence contours for models with two MG parameters at $z=0$. On the left hand side we present the constraints for ($\mu$, $\eta$) while on the right hand side we show the results for ($\mu$, $\Sigma$), for various combinations of datasets in each case (the dataset acronyms are provided in Table \ref{Table:DataSets}). The dashed lines correspond to the GR values for each parameter. Here, we find that P18 data has an above 3-$\sigma$ tension with respect to GR, while adding SNe+BAO+RSD worsens the tension.}
\label{Fig:2MG_constraints}
\end{figure*}
\section{Datasets \label{sec:datasets}}
We make use of various current cosmological datasets that are relevant to our analysis. A significant contribution to the constraints on the cosmological parameters comes from the latest Planck mission 2018 measurements of the CMB temperature and polarization, which are already implemented in the current version of \texttt{ISiTGR}. At high-$\ell$ Planck provides the temperature auto correlation for $30 \leq\ell\leq 2058$, the E-mode polarization auto correlation at $30 \leq\ell\leq 1996$, and the cross correlation between temperature and E-mode in the range $30 \leq\ell\leq 1996$ \cite{Planck2018}. Here, we use the combined temperature and polarization spectra at $\ell\geq 30$ together with the low-$\ell$ temperature Commander likelihood and we label this joint dataset as TTTEEE. For the region $2 \leq\ell\leq 29$ we employ the SimAll likelihood code for EE spectra \footnote{Only the EE likelihood is used due to a poor statistical consistency of the TE spectrum that was reported in \cite{Planck2018}} and we refer to it as lowE. Additionally, we make use of the gravitational lensing measurements from the CMB provided by the Planck collaboration \cite{Planck-2018-lensing}.

In order to obtain tighter constraints and break degeneracies between parameters we use the Pantheon sample data presented in \cite{Pan-STARRS-2017}, which combines 279 Supernovae type Ia (SNe) with redshift range $(0.03 < z < 0.68)$ from the Sloan Digital Sky Survey (SDSS), Supernova Legacy Survey (SNLS), and various low-z and HST samples, giving a total of 1048 SNe ranging from $(0.01 < z < 2.3)$. We combine this data with Baryon Acoustic Oscillations (BAO) and Redshift Space Distortions (RSD) measurements from the Baryon Oscillation Spectroscopic Survey (BOSS) Data Release 12 (DR12) which combines individual results from different works into a set of consensus values and likelihoods in three effective redshifts, being 0.38, 0.51 and 0.61 \cite{AlamEtAl2016}. Furthermore, we use BAO measurements from the Six Degree Field Galaxy Survey (6dFGS) at $z_{\text{eff}}=0.106$ \cite{BeutlerEtAl2011} and the SDSS Data Release 7 Main Galaxy Sample (MGS) at $z_{\text{eff}}=0.15$ \cite{RossEtAl2014}. Moreover, we consider the quasar-Lyman-$\alpha$ cross-correlation combined with Lyman-$\alpha$ forest auto-correlation measurements at $z_{\text{eff}}=2.35$ \cite{2019-BAO-lyalpha-quasar-Blomqvist}. Additionally, we use Big Bang Nucleosynthesis (BBN) data from the primordial deuterium abundance $D/H=2.527 \pm 0.030 \times 10^{-5}$ \cite{Cooke2018} and local measurements of the Hubble constant $H_0$ from the Hubble Space Telescope (HST) \cite{Riess2019H0}.

We further consider the galaxy clustering and weak lensing data from the Dark Energy Survey (DES) Year 1 analysis \cite{DES2017}. However, since the MG approach followed in this work does not support a non-linear prescription, we only consider data points that correspond to linear scales following a similar procedure to other works \cite{Zucca:2019xhg,DESMG2018,Planck2015MG}. More precisely, we define $\Delta\chi^2 = (\textbf{d}_{\text{NL}}-\textbf{d}_{\text{L}})^\text{T} \textbf{C}^{-1} (\textbf{d}_{\text{NL}}-\textbf{d}_{\text{L}})$ where $\textbf{C}$ is the full DES-Y1 covariance matrix, and $\textbf{d}_{\text{NL}}$ and $\textbf{d}_{\text{NL}}$ are vectors of the form $(\xi_\pm, \gamma_T, w_\theta)$, calculated using the non-linear and linear matter power spectrum, respectively. This definition of $\Delta\chi^2$ quantifies how well the $3\times 2$ pt correlation functions using the non-linear and linear theory fit each other. We calculate $\Delta\chi^2$ in the standard $\Lambda$CDM best-fit model and then we identify which data point contributes the most to $\Delta\chi^2$. Next, we remove this data point and continue this process recursively until we get $\Delta\chi^2<1$. Therefore, this procedure removes the data points corresponding to scales where the non-linear and linear regimes disagree, so that $\Delta\chi^2<1$. 

We limit ourselves to combining datasets that are in good agreement based on previous analysis \cite{IOI2,GQC2019}. However, we still combine some datasets that can lead to mild tensions in $\sigma_8$ such as Planck and DES, but we avoid to use datasets where this tension with Planck is higher such as the Kilo-Degree Survey data. Additionally, we discuss how this tension behaves in the context of MG in section 4.1. Finally, we summarize the individual datasets and combinations that we use in this work in Table \ref{Table:DataSets}. 
\begin{figure*}
\begin{tabular}{c c}
{\includegraphics[width=8.2cm]{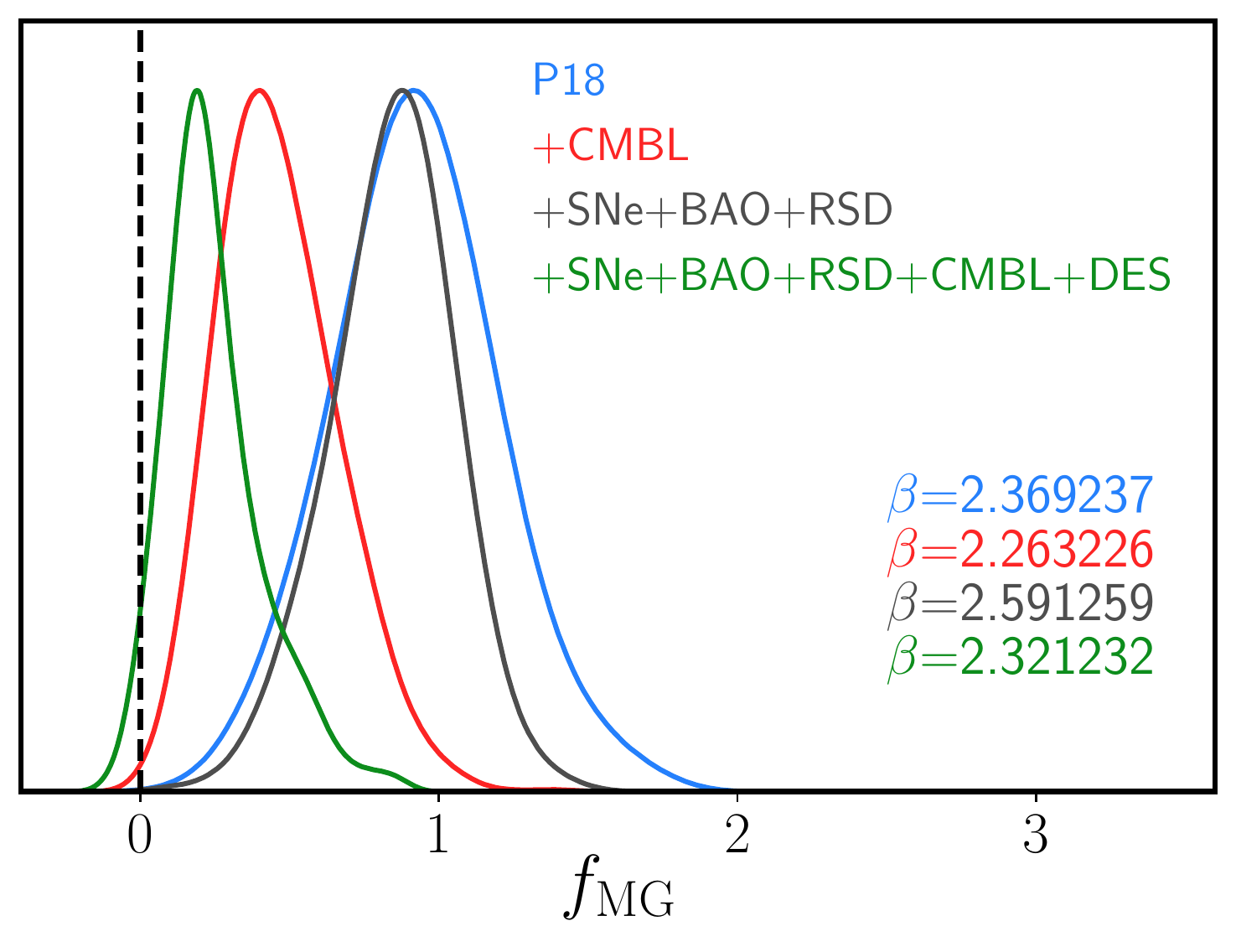}} &
{\includegraphics[width=8.2cm]{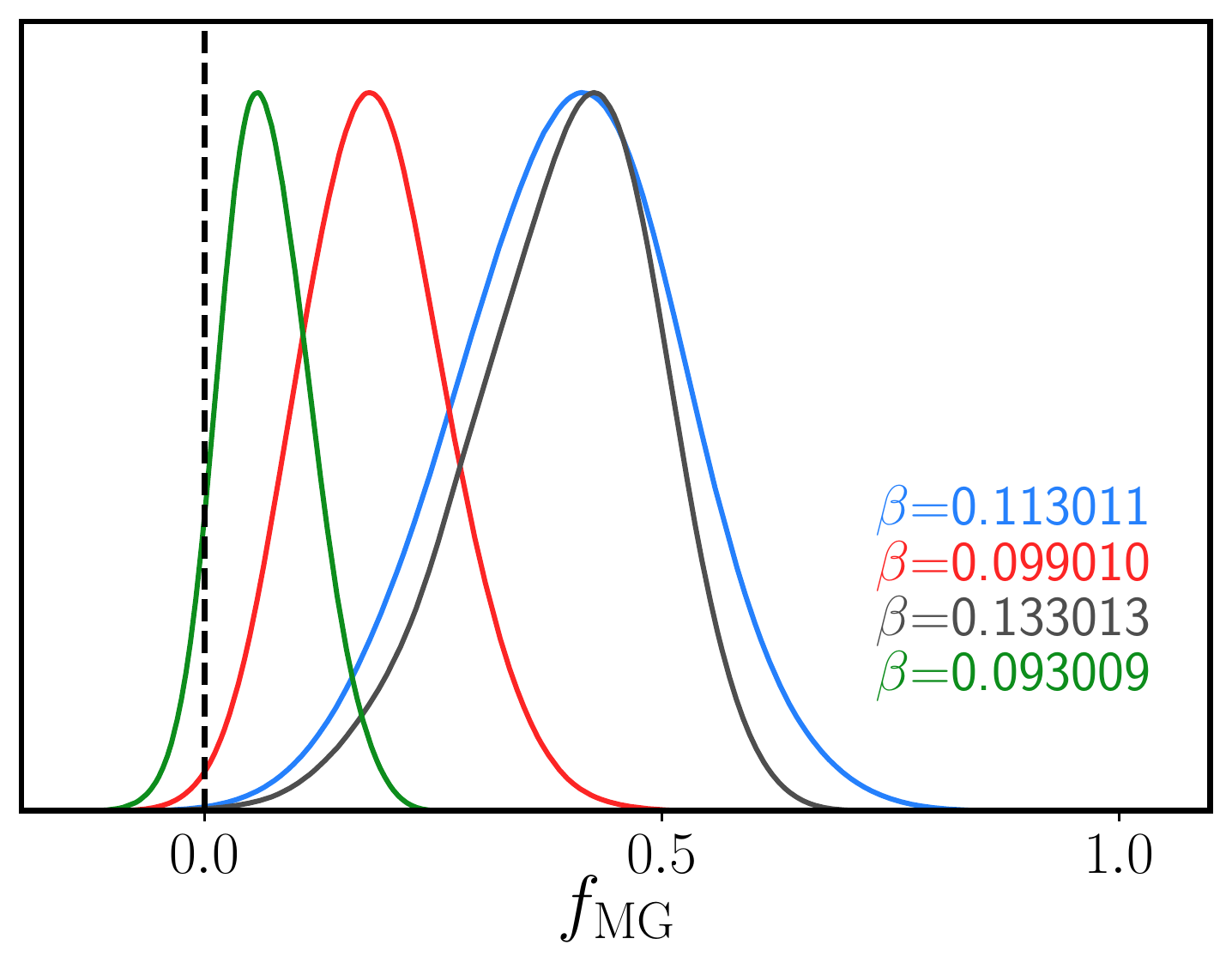}}
\end{tabular}
\caption{Associated probability distribution functions of the $f_{\text{MG}}$ parameter for each of the constraints shown in Fig. \ref{Fig:2MG_constraints}. The left panel corresponds to constraints for ($\mu$, $\eta$) while the right panel is associated with ($\mu$, $\Sigma$). We list the corresponding values of $\beta$ for each distribution. Also, $f_{\text{MG}}=0$ represents the corresponding GR value for the MG parameters.}
\label{Fig:2MG_constraints_fMG}
\end{figure*}
\begin{table}
\begin{center}
\scriptsize 
 \begin{tabular} { c |  >{\centering}m{1.1cm} |  >{\centering}m{1.4cm} |  >{\centering}m{1.1cm} | c }
\hline
& \multicolumn{2}{ c| }{Planck collaboration results} & \multicolumn{2}{ c }{\texttt{ISiTGR} pipeline} \\ \cline{2-5}
parameter & P18 & P18+CMBL & P18 & P18+CMBL \\ \hline
$\mu_0 -1$ & $0.12^{+0.29}_{-0.51}$ & $0.10^{+0.30}_{-0.42}$ & $0.12^{+0.28}_{-0.54}$ & $0.09^{+0.27}_{-0.44}$ \\

$\eta_0 -1$ & $0.55^{+0.78}_{-0.1.2}$ & $0.22^{+0.55}_{-1.0}$ & $0.65^{+0.83}_{-1.3}$ & $0.27^{+0.56}_{-1.0}$ \\

$\Sigma_0 -1$ & $0.27^{+0.15}_{-0.13}$ & $0.100 \pm 0.093$ & $0.29^{+0.15}_{-0.13}$ & $0.106\pm 0.092$ \\
\hline
\end{tabular}
\end{center}
\caption{Comparison of the constraints for MG parameters obtained using our own pipeline versus the original results presented in \citet{Planck2018} using the Planck 2018 data. This shows consistency between the two pipelines for the study of models with two MG parameters.}
\label{Table:Validation}
\end{table}

\begin{figure}
 \centering
{\includegraphics[width=8cm]{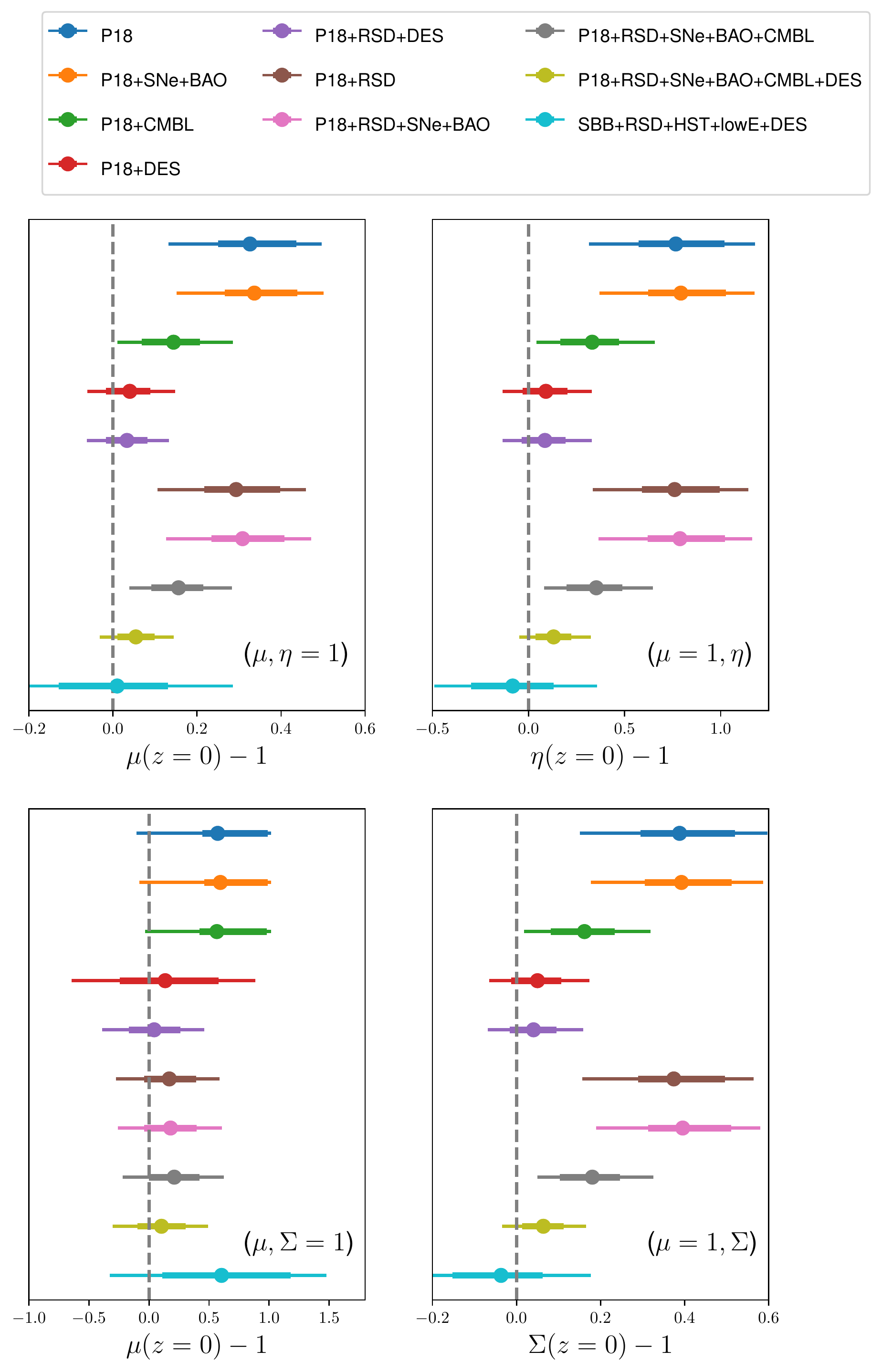}}
\caption{Expectation value, 68\% (thin error bars) and 95\% (thick error bars) confidence limits for each MG parameter given the models shown in Table \ref{Table:MG-models}. The different colors represent the datasets used for the constraints and the dashed vertical line represents the MG parameter values that correspond to GR.}
\label{Fig:1MG_constraints}
\end{figure}
\begin{figure}
{\includegraphics[width=8cm]{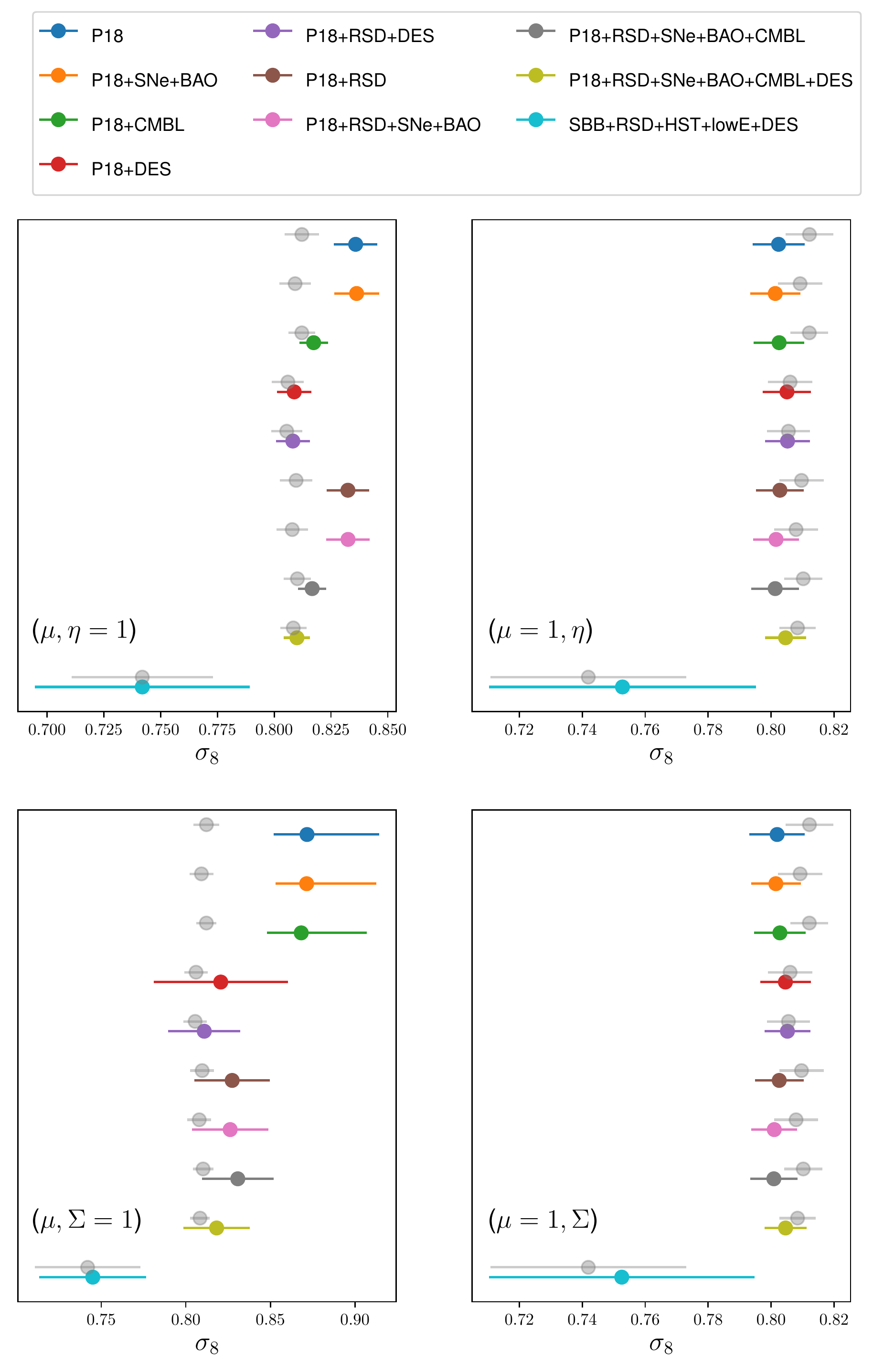}}
\caption{Expectation value and 68\% confidence interval for the $\sigma_8$ parameter for four MG models and the $\Lambda$CDM model. The different colors represent the datasets used for the constraints and are in the same order as in Fig. \ref{Fig:1MG_constraints}. we also plot the corresponding $\Lambda$CDM constraints on $\sigma_8$ in gray for comparison. We observe that for models $(\mu=1,\eta)$ and $(\mu=1,\Sigma)$, the measurements on $\sigma_8$ are in slightly better agreement than those of $\Lambda$CDM.}
\label{Fig:1MG_sigma8}
\end{figure}
\section{Results}
In this section we present our results for constraints on MG parameters. We obtain the constraints for these models by using a Markov Chain Monte Carlo (MCMC) sampling process where we vary the MG parameters in addition to the six core cosmological parameters. Namely, the baryon physical density parameter $\Omega_b h^2$, the cold dark matter physical density $\Omega_c h^2$, the angular size of the sound horizon $\theta$, the reionization optical depth $\tau$, the spectral index $n_s$ and $\ln(10^{10} A_s)$, the amplitude of the primordial power spectrum. The different MG parameter combinations we use and the relationships between them in each case are shown in Table \ref{Table:MG-models}, while priors for the parameters can be found in Table \ref{Table:Priors}. In addition to the cosmological and MG parameters, we also varied the nuisance parameters associated to DES clustering and weak lensing data as in \cite{DES2017} and Planck 2018 calibration parameters (see Table 16 in \cite{Planck18-likelihoods}). We also varied the absolute magnitude parameter for SNe but this is internally marginalized within the SNe likelihood implemented in \texttt{CosmoMC} by using the analytic marginalization procedure described in \cite{ConleyEtAl2011}. We use \texttt{CosmoMC} \cite{COSMOMC} as MCMC sampler with the addition of the MG parameters. Also, as a general rule, all our set of chains have a convergence of $R-1 \lesssim 0.08$ based on the Gelman-Rubin statistic \cite{2013-Lewis-EfficientSampling,Gelman-Rubin} and we fix the sum of neutrino masses to 0.060 eV with a normal mass hierarchy. Importantly, besides the case in which we consider two MG parameters, we are considering models where only one MG parameter is varied while the other is kept at its GR value. Therefore, we assume there is either no gravitational slip effect, or there are no departures from GR in the growth rate of structure, or there are no deviations in the motion of light from what GR predicts. 

Regardless of the equivalence between models such as ($\mu$, $\eta$) and ($\mu$, $\Sigma$), we opt to perform an independent analysis in case some differences may arise from the point of view of MCMC parameter space sampling. Also, we compare with the models presented in \cite{Linder2020-limitedMG}, and for that we present some relationships between the nomenclature used therein and the MG parameters used here: in a matter-dominated epoch and with negligible anisotropic stress we have that our MG parameters can be expressed as $\mu=G_{\text{matter}}/G$ and $\Sigma=G_{\text{light}}/G$. However, the slip parameter presented in that work is related to ours as $\bar{\eta}=2/(1+\eta)$.

In the following, we show the corresponding constraints and analysis for these MG models and quantify the level of tension with respect to GR. Later on, we perform a model comparison through a Bayesian analysis for different datasets and provide an interpretation of the results.

\subsection{MG constraints}
We search for deviations from GR by considering the parameter set $\theta_{\text{base}}=\{ \Omega_b h^2, \Omega_c h^2, \theta, \tau, n_s, A_s, X_1, X_2 \}$, where $X_1$ represents $E_{11}$ or $\mu_0$, and $X_2$ corresponds to either $E_{22}$ or $\Sigma_0$, depending on the parameterization we are using. However, we set $X_1=0$ or $X_2=0$ when we consider models with only one varying MG parameter.

In order to validate our pipeline we perform a basic comparison check, varying two MG parameters. We compare the constraints obtained using \texttt{ISiTGR} with the latest results from the Planck collaboration using the Planck 2018 data. We show in Table \ref{Table:Validation} the comparison of the constraints in the three MG parameters. We find that the results for two MG parameters are in good agreement when we use the Planck data.

In the case of models with two varying MG parameters we quantify the tension by considering a linear combination of the MG parameters. As argued in \cite{Planck2015MG}, this linear combination corresponds to the degeneracy line that approximately allows the maximum departure from GR in the MG parameter space. Namely, we define the function
\begin{equation}
f_{\text{MG}} = \beta[\mu(z=0)-1]+[X(z=0)-1],
\label{eq:fmg}
\end{equation}
where $\beta$ is a coefficient that is related to the slope of the degeneracy line which maximizes the tension with respect to GR, and $X$ is equal to $\eta$ or $\Sigma$ depending on which model we are considering. We can observe that $f_{\text{MG}}=0$ for the GR case and it can be evaluated by using equations (\ref{eq:explicit-form_mueta1}) and (\ref{eq:explicit-form_mueta2}), or (\ref{eq:explicit-form_muSigma1}) and (\ref{eq:explicit-form_muSigma2}). We proceed to find the $\beta$ coefficient in each case by a simple numerical procedure which approximately returns the strongest allowed tension with respect to GR. We show the constraints for some datasets given the ($\mu$, $\eta$) and ($\mu$, $\Sigma$) models in the two-dimensional MG parameter space in Fig. \ref{Fig:2MG_constraints}, while the corresponding results for $f_{\text{MG}}$ can be found in Fig. \ref{Fig:2MG_constraints_fMG}. One of the positive aspects of using (\ref{eq:fmg}) is that we find that even though the individual distributions of $\mu(z=0)-1$, $\eta(z=0)-1$ and $\Sigma(z=0)-1$ may not be Gaussian, the linear combination given by $f_{\text{MG}}$ produces a better approximation to a Gaussian distribution than the posterior distribution of some MG parameters, as we verified by looking at quantile-quantile plots. Thus, it gives a more reliable way to quantify the tension by computing its $\chi^2$ in one dimension. 

As we can see in Fig. \ref{Fig:2MG_constraints}, for both ($\mu$, $\eta$) and ($\mu$, $\Sigma$) parameterizations, the most significant departure from GR is obtained when using P18+SNe+BAO+RSD. While there is no clear sign which combination of MG parameters is responsible for the tension when using ($\mu$, $\eta$), we can observe that in the ($\mu$, $\Sigma$) parameter space at $z=0$, the tension with respect to GR can be particularly attributed to $\Sigma$. Then, the tension with GR can be reduced by adding gravitational lensing data to P18+SNe+BAO+RSD such as CMBL, or it can be fully removed by adding DES. The fact that adding lensing data shifts the contours towards GR when combined with P18 is related to correlations in the parameter space. The lensing amplitude is proportional to the product $\Sigma_0 S_8$ and these two parameters are anti-correlated. Then, the fact that CMB data such as P18 prefers a high value of $S_8$ compared to DES, makes that P18+DES favours a lower value of $\Sigma_0$. The same effect occurs to the lensing amplitude for the lensing of the CMB but to a lower degree, since CMBL prefers a value of S8 higher than that of DES. Therefore, DES reduces more the tension with respect to GR than CMBL, when combined with P18.

For models that consist of one free MG parameter, we obtain Gaussian constraints in most of the cases \footnote{We obtain three cases in Fig. \ref{Fig:1MG_constraints} where constraints are clearly non-Gaussian, due to the lack of relevant data to properly constrain $\mu$.}. We show in Fig. \ref{Fig:1MG_constraints} the constraints for each model with a single MG parameter using several datasets. We find that there is a tension when using Planck data without lensing, except for ($\mu$, $\Sigma=1$) where we find that the tension with respect to GR is not significant. However, when we vary $\Sigma$ but set either $\mu$ or $\eta$ to its GR value, we observe that there is a similar trend for these models: there is a 3.4-$\sigma$ to 3.5-$\sigma$ departure from GR when using Planck data alone with no lensing. This tension can increase if we add probes such as SNe, BAO and RSD on top of P18. We point out that the MG models with one free parameter provide a different scenario from models varying two MG parameters at the same time. While deviations with respect to GR manifest only in $\Sigma$ if we look to models such as ($\mu$, $\Sigma$), models with one MG parameter show that the departures from GR are actually occurring not only in $\Sigma$, but also in either the growth or in the gravitational slip. Indeed, as we show in Table \ref{Table:MG-models}, for all models with one MG parameter there is a one-to-one correspondence between the two remaining MG parameters. Then, if we consider all the models with one MG parameter except for ($\mu$, $\Sigma=1$), this one-to-one correspondence is the reason why even though the lensing data has a constraining effect on $\Sigma$, the error bars in $\mu$ shrink and the value of $\mu$ is also displaced towards its GR value in the upper left panel in Fig. \ref{Fig:1MG_constraints}. Hence, we also find that lensing data reduces the tension with respect to GR in all the cases. While CMBL alleviates the tension produced by P18, DES is able to restore GR, as was observed in the case of models with two MG parameters.

In Table \ref{Table:tension_functional}, we show the  level of tension with respect to GR for the models given in Table \ref{Table:MG-models} using the $n$-$\sigma$ level of tension and the Index of Inconsistency (IOI) as given in e.g. \cite{IOI1,GQC2019}. We can observe that while P18 has about a 3.2-$\sigma$ tension with GR for models with two MG parameters, this tension can go up to 3.5-$\sigma$ if we do not allow gravitational slip ($\eta=1$). Adding other datasets such as SNe, BAO and RSD to P18 leads to a higher tension, reaching a maximum of 3.9-$\sigma$. 
However, the tension gets significantly reduced when including CMBL and practically goes away when adding DES, as we can observe in combinations such as P18+RSD+CMBL and P18+RSD+DES. We note that we need to add RSD since it provides better constraints on $\mu$ rather than using P18+CMBL or P18+DES alone, which can lead to some marginal tension values due mostly to contour volume effects. To confirm such effects, we produce simulated CMB Planck-like data for the $\Lambda$CDM model and analyze it through our pipeline \footnote{We use the python script makePerfectForecastDataset.py that is present in \texttt{CosmoMC} with theoretical $C_{\ell}$ produced by \texttt{ISiTGR}}. We found that the MG parameters mean values are in agreement with GR. Due to these volume effects we observed a tension of 0.1-$\sigma$ with GR for $\mu$. However, we found that these effects are quite negligible in terms of deviations from GR.

\begin{table*}
\begin{center}
 \begin{tabular} { p{1.5cm}  | >{\centering}m{0.5cm}| >{\centering}m{0.5cm}| >{\centering}m{0.5cm} |>{\centering}m{0.5cm}| >{\centering}m{0.5cm}| >{\centering}m{0.5cm} |>{\centering}m{0.5cm}| >{\centering}m{0.5cm}| >{\centering}m{0.5cm} |>{\centering}m{0.5cm}| >{\centering}m{0.5cm}| >{\centering}m{0.5cm}| >{\centering}m{0.5cm}| >{\centering}m{0.5cm}| >{\raggedleft\arraybackslash}m{0.6cm}}
\hline
 Model & \rotatebox{90}{\textbf{P18}} & \rotatebox{90}{\textbf{P18+SNe+BAO}} & \rotatebox{90}{P18+CMBL} & \rotatebox{90}{P18+DES} & \rotatebox{90}{P18+CMBL+DES} & \rotatebox{90}{\textbf{P18+RSD}} & \rotatebox{90}{P18+RSD+CMBL} & \rotatebox{90}{P18+RSD+DES} & \rotatebox{90}{P18+RSD+CMBL+DES} & \rotatebox{90}{\textbf{P18+SNe+BAO+RSD}} & \rotatebox{90}{P18+SNe+BAO+RSD+CMBL} & \rotatebox{90}{P18+SNe+BAO+RSD+DES} & \rotatebox{90}{P18+SNe+BAO+RSD+CMBL+DES}  & \rotatebox{90}{SBB+RSD+HST+lowE} & \rotatebox{90}{SBB+RSD+HST+lowE+DES} \\
\hline
$(\mu,\eta)$ & \textbf{3.2} & \textbf{3.4} & 2.2 & 1.5 & 1.5 & \textbf{3.4} & 2.4 & 1.2 & 1.4 & \textbf{3.9} & 2.6 & 1.2 & 1.4 & 1.8 & 1.4 \\
$(\mu,\Sigma)$ & \textbf{3.2} & \textbf{3.6} & 2.3 & 1.5 & 1.9 & \textbf{3.5} & 2.3 & 0.8 & 1.1  & \textbf{3.8} & 2.5 & 1.0 & 1.4 & 2.1 & 1.7 \\
$(\mu,\eta=1)$ & \textbf{3.5} & \textbf{3.8} & 2.1 & 0.8 & 1.1 & \textbf{3.2} & 2.3 & 0.7 & 1.1 & \textbf{3.5} & 2.5 & 0.9 & 1.2 & 1.6 & 0.1 \\
$(\mu=1,\eta)$ & \textbf{3.5} & \textbf{3.9} & 2.1 & 0.8 & 1.0 & \textbf{3.7} & 2.3 & 0.7 & 1.1  & \textbf{3.9} & 2.4 & 0.9 & 1.4 & 0.7 & 0.4 \\
$(\mu=1,\Sigma)$ & \textbf{3.4} & \textbf{3.7} & 2.1 & 0.8 & 1.2 & \textbf{3.6} & 2.3 & 0.7 & 1.1  & \textbf{3.9} & 2.5 & 0.9 & 1.3 & 0.4 & 0.3 \\
$(\mu,\Sigma=1)$ & 1.8 & 1.9 & 1.8 & 0.3 & 0.6 & 0.8 & 0.9 & 0.2 & 0.4 & 0.8 & 1.0 & 0.3 & 0.5 & 1.1 & 1.2 \\
\hline
\end{tabular}
\end{center}
\caption{Typical $n$-$\sigma$ level of tension with respect to GR for the MG models shown in Table \ref{Table:MG-models} using different datasets. We write in bold the $n$-$\sigma$ values and datasets that produce a tension above 3-$\sigma$ for some MG model.}
\label{Table:tension_functional}
\end{table*}

\begin{table*}
\begin{center}
 \begin{tabular} { p{1.5cm}  | >{\centering}m{0.5cm}| >{\centering}m{0.5cm}| >{\centering}m{0.5cm}| >{\centering}m{0.5cm}| >{\centering}m{0.5cm} |>{\centering}m{0.5cm}| >{\centering}m{0.5cm}| >{\centering}m{0.5cm}| >{\centering}m{0.5cm} |>{\centering}m{0.5cm}| >{\centering}m{0.5cm} |>{\centering}m{0.5cm} | >{\centering}m{0.5cm}| >{\centering}m{0.5cm}| >{\raggedleft\arraybackslash}m{0.6cm}}
\hline 
 Model & \rotatebox{90}{\textbf{P18}} & \rotatebox{90}{\textbf{P18+SNe+BAO}} & \rotatebox{90}{P18+CMBL} & \rotatebox{90}{P18+DES} & \rotatebox{90}{P18+CMBL+DES} & \rotatebox{90}{\textbf{P18+RSD}} & \rotatebox{90}{P18+RSD+CMBL} & \rotatebox{90}{P18+RSD+DES} & \rotatebox{90}{P18+RSD+CMBL+DES} & \rotatebox{90}{\textbf{P18+SNe+BAO+RSD}} & \rotatebox{90}{P18+SNe+BAO+RSD+CMBL} & \rotatebox{90}{P18+SNe+BAO+RSD+DES} & \rotatebox{90}{P18+SNe+BAO+RSD+CMBL+DES}  & \rotatebox{90}{SBB+RSD+HST+lowE} & \rotatebox{90}{SBB+RSD+HST+lowE+DES} \\
\hline
$(\mu,\eta)$ & \textbf{5.0} & \textbf{5.8} & 2.4 & 1.1 & 1.2 & \textbf{5.9} & 2.9 & 0.7 & 0.9 & \textbf{7.6} & 3.3 & 0.7 & 1. & 1.6  & 1.0 \\
$(\mu,\Sigma)$ & \textbf{5.0} & \textbf{6.3} & 2.6 & 1.1 & 1.7 & \textbf{6.0} & 2.6 & 0.3  & 0.6 & \textbf{7.0} & 3.2 & 0.5 & 0.9 & 2.1 & 1.5 \\
$(\mu,\eta=1)$ & \textbf{6.1} & \textbf{7.1} & 2.1 & 0.3 & 0.6 & \textbf{5.2} & 2.6 & 0.2 & 0.6 & \textbf{6.2} & 3.2  & 0.4 & 0.7 & 1.3 & 0.0 \\
$(\mu=1,\eta)$ & \textbf{6.1} & \textbf{7.4} & 2.3 & 0.3 & 0.5 & \textbf{6.8} & 2.6 & 0.3 & 0.6 & \textbf{7.5} & 3.0 & 0.4 & 0.9 & 0.3 & 0.1 \\
$(\mu=1,\Sigma)$ & \textbf{5.6} & \textbf{6.8} & 2.2  & 0.3 & 0.7 & \textbf{6.3}  & 2.7 & 0.3 & 0.6 & \textbf{7.6} & 3.2 & 0.4 & 0.8 & 0.1 & 0.1 \\
$(\mu,\Sigma=1)$ & 1.5 & 1.8 & 1.7 & 0.1 & 0.2 & 0.3  & 0.4 & 0.0 & 0.1  & 0.3 & 0.5 & 0.0 & 0.1 & 0.6 & 0.7 \\
\hline
\end{tabular}
\end{center}
\caption{Analogous to Table \ref{Table:tension_functional} but in terms of the IOI values. Here we can interpret a value of IOI$<1$ as no significant inconsistency with respect to GR, $1<$IOI$<2.5$ indicates a weak inconsistency, $2.5<$IOI$<5$ means a moderate inconsistency, while an IOI$>5$ stands for a strong inconsistency. Hence, we list in bold the datasets and IOI values that represent a strong inconsistency with respect to GR.}
\label{Table:tension_IOI}
\end{table*}

\begin{table*}
\begin{center}
 \begin{tabular} {  p{1.3cm}  | >{\centering}m{0.59cm} | >{\centering}m{0.59cm} | >{\centering}m{0.59cm}| >{\centering}m{0.59cm}| >{\centering}m{0.59cm}| >{\centering}m{0.59cm} |>{\centering}m{0.59cm} |>{\centering}m{0.59cm}| >{\centering}m{0.59cm}| >{\centering}m{0.59cm} |>{\centering}m{0.59cm}| >{\centering}m{0.59cm} |>{\centering}m{0.59cm}| >{\centering}m{0.59cm} | >{\raggedleft\arraybackslash}m{0.6cm}}
\hline 
 $\ln B_{i0}$ & \rotatebox{90}{\textbf{P18}} & \rotatebox{90}{\textbf{P18+SNe+BAO}} & \rotatebox{90}{P18+CMBL} & \rotatebox{90}{P18+DES} & \rotatebox{90}{P18+CMBL+DES} & \rotatebox{90}{\textbf{P18+RSD}} & \rotatebox{90}{P18+RSD+CMBL} & \rotatebox{90}{P18+RSD+DES} & \rotatebox{90}{P18+RSD+CMBL+DES}  & \rotatebox{90}{\textbf{P18+SNe+BAO+RSD}} & \rotatebox{90}{P18+SNe+BAO+RSD+CMBL} & \rotatebox{90}{P18+SNe+BAO+RSD+DES} & \rotatebox{90}{P18+SNe+BAO+RSD+CMBL+DES} &  \rotatebox{90}{SBB+RSD+HST+lowE} & \rotatebox{90}{SBB+RSD+HST+lowE+DES} \\
\hline
$(\mu, \eta)$ & -0.1 &  0.5 & -2.8 & -5.1 & -4.2 & -0.8 & -3.2 & -6.2 & -5.4 & -0.3 & -2.7 & -5.8 & -5.5 & -1.0 & -4.0 \\
$(\mu, \Sigma)$ & 0.3 &  0.8 & -2.1 & -4.9 & -3.8 & -1.1 & -3.7 & -7.1 & -6.1 & -0.5  & -3.1 & -7.2 & -5.7 & -1.2 & -3.9 \\
$(\mu, \eta=1)$ & 0.8 &  \textbf{1.6} & -1.5  & -4.4 & -3.2 &  0.4 & -1.2  & -4.0 & -4.2 & 0.7 & -0.8 & -5.0 & -3.4 & -0.2 & -3.5 \\
$(\mu=1, \eta)$ & \textbf{1.8} &  \textbf{2.5} & -0.8 & -3.7 & -2.4 &  \textbf{2.0} & -0.4 & -3.9 & -3.3 & \textbf{2.6} &  0.1 & -3.7 & -2.1 & -0.6 & -3.0 \\
$(\mu=1, \Sigma)$ & \textbf{1.3} &  \textbf{1.8} & -1.4 & -4.8 & -3.3 &  \textbf{1.7} & -1.0 & -1.7 & -3.3 & \textbf{1.9} & -0.5 & -4.5 & -3.4 & -0.8 & -3.4 \\
$(\mu, \Sigma=1)$ & -1.4 & -1.2 & -0.8 & -2.6 & -1.8 & -3.1 & -2.1  & -3.4  & -2.9 & -3.2 & -2.4 & -3.9 & -2.8 & -0.9 & -0.7 \\
\hline
\end{tabular}
\end{center}
\caption{Results for the natural logarithm of the Bayes factor with respect to $\Lambda$CDM for different MG parameterizations. We show in bold the datasets in which $\Lambda$CDM is not the favoured model by the data.  In each column, the values written in bold mean that there is a moderate preference for the MG model over $\Lambda$CDM. See Table \ref{Table:JeffreyScale} for the interpretation of the Bayes factor values.}
\label{Table:BayesFactor}
\end{table*}
\begin{table}
\begin{center}
\begin{tabular} { >{\centering}p{2.5cm} | >{\centering\arraybackslash}p{4.5cm} }
\hline
 Ranges & Interpretation \\ \hline 
 $0<|\ln B_{i0}|<1$ & Not worth more than a bare mention \\ 
 $1\leq|\ln B_{i0}|<3$ & Moderate  \\ 
 $3\leq|\ln B_{i0}|<5$ & Strong  \\  
 $5\leq|\ln B_{i0}|$ & Very strong \\ 
\hline
\end{tabular}
\caption{Adopted Jeffrey's scale for the interpretation of the Bayes factor. We use a similar Jeffrey's scale as presented in \citet{10.2307/2291091,2017PhRvL.119j1301H}.}
\label{Table:JeffreyScale} 
\end{center}
\end{table}
\begin{table}
\begin{center}
\begin{tabular} { >{\centering}p{2.5cm} | >{\centering\arraybackslash}p{3cm} }
\hline
Parameter & Prior range \\ \hline
$\Omega_b h^2$ & $[0.005, 0.1]$ \\
$\Omega_c h^2$ & $[0.001, 0.99]$ \\
$\theta$ & $[0.5, 10]$ \\
$\tau$ & $[0.01, 0.8]$ \\
${\rm{ln}}(10^{10} A_s)$ & $[1.61, 3.91]$ \\
$n_s$ & $[0.8, 1.2]$ \\ \hline
$E_{11}$ & $[-5, 5]$ \\
$E_{22}$ & $[-5, 5]$ \\
$\mu_0$ & $[-3, 3]$ \\
$\Sigma_0$ & $[-3, 3]$ \\
\hline
\end{tabular}
\caption{Parameters and prior ranges used in the analysis. For models with one MG parameter, we set some of the free MG parameters equal to zero. For example in ($\mu=1$, $\eta$) we set $E_{11}=0$ but keep the prior range for $E_{22}$, and so on for the other MG models.}
\label{Table:Priors} 
\end{center}
\end{table}
While the combination SBB+RSD+HST+lowE does not constrain the MG parameters very well, SBB+RSD+HST+lowE+DES does not give a precise measure of $\eta_0$ but shows some tension in $\mu_0$. Therefore, setting $\mu=1$ leads to smaller $n$-$\sigma$ values for models with one free MG parameter compared with models with two varying MG parameters.

It is important to recall that we are fixing the neutrino mass in our analysis of tensions. It is well known that massive neutrinos can suppress the growth of structure at small scales, while $\mu$ can produce a similar effect or enhancement. However, previous works modeling scale-independent parameterizations while varying the neutrino mass showed no significant correlation in such a case \cite{Zucca:2019xhg, ISiTGR-CGQ}. Therefore, we do not expect a significant correlation between the MG parameters and the neutrino mass while using a scale-independent parameterization. On the other hand, adding another free parameter will produce an increase in the contours for the MG parameters due to marginalization, probably leading to weaker but still significant tensions.

Alternatively, we use the Index of Inconsistency (IOI) \cite{IOI1,GQC2019} in order to quantify the tension with respect to GR with the knowledge that in one dimension the relationship  $n$-$\sigma=\sqrt{2\text{IOI}}$ is satisfied. Therefore, we can use the values of Table \ref{Table:tension_functional} to calculate the degree of inconsistency with respect to GR in terms of IOI. This applies thus for cases where we vary a single MG parameter. However, since IOI requires distribution functions that are nearly Gaussian, we opt to calculate IOI using the distribution of $f_{\text{MG}}$ in the cases where we vary two MG parameters. Hence, we can still convert the $n$-$\sigma$ value to IOI for cases where we use two MG parameters, since $n$-$\sigma$ is computed for a single parameter $f_{\text{MG}}$. IOI is a moment-based measure that ideally uses the full-parameter space to quantify the consistency between two datasets, assuming Gaussian distributions. However, in terms of consistency with respect to GR we limit ourselves to use the marginalized posterior distribution of $f_{\text{MG}}$, since is this function what quantifies deviations from GR. We do not expect the size of the flat prior adopted for the MG parameters having any dependency over the calculated IOI.

We show in Table \ref{Table:tension_IOI} the corresponding values of IOI and interpret them using a calibrated version of the Jeffrey's scale for IOI. We can observe a similar trend from comparison of Tables \ref{Table:tension_functional} and \ref{Table:tension_IOI}: the datasets P18, P18+SNe+BAO, P18+RSD and P18+SNe+BAO+RSD seem to show a strong inconsistency with GR. Furthermore, while IOI applied to the full Planck data with lensing generally indicates a weak inconsistency with respect to GR, P18+CMBL, along with other probes such as SNe, BAO and RSD, seem to point to a moderate inconsistency for most of the models. Also, Planck data in combination with DES and other probes (for example P18+RSD+DES) shows no significant inconsistency.

Additionally, we notice that looking at the improvement in $\chi^2$ of MG over GR among the MG models where one MG parameter is fixed to its GR value, only model $(\mu,\Sigma=1)$ is not able to significantly improve the fit over $\Lambda$CDM. The improvement in $\chi^2$ also reveals that models $(\mu=1,\Sigma)$ and $(\mu=1,\eta)$ fit the data slightly better than $(\mu,\eta=1)$ when P18 is combined with probes that does not include lensing. This comparison between the MG models using the improvement in $\chi^2$ leads to similar conclusion if we look at IOI or the Bayes factor as shown in section \ref{sec:model-comparison}. Now, it is important to mention that significant deviations from GR in one of these specific models might point out to limiting modified gravity models of a more general theory, as suggested in \cite{Linder2020-limitedMG} for the Horndeski class.

Finally, due to correlations with MG parameters, we analyze the constraints from Planck and cosmic shear experiments on the amplitude of matter density fluctuations, usually parameterized by the $\sigma_8$ parameter. Hence, we show the constraints on $\sigma_8$ for models with one MG parameter as well as for the  $\Lambda$CDM model in Fig. \ref{Fig:1MG_sigma8}. For $\Lambda$CDM, we observe that there is some tension between P18 and SBB+RSD+HST+lowE+DES, at about the 2.2-$\sigma$ level. A comparison of Fig. \ref{Fig:1MG_constraints} and Fig. \ref{Fig:1MG_sigma8} shows that for both ($\mu$, $\eta=1$) and ($\mu$, $\Sigma=1$) models there is a positive correlation between $\mu$ and $\sigma_8$ when P18 is used. This leads to a higher tension in $\sigma_8$ as compared to the $\Lambda$CDM model. Thus, although ($\mu$, $\eta=1$) shows a higher departure from GR than ($\mu$, $\Sigma=1$), it does not solve the tension in $\sigma_8$. As we mentioned before, besides the MCMC parameter space sampling, ($\mu=1$, $\eta$) and ($\mu=1$, $\Sigma$) represent the same model. Now, if we set $\mu=1$, then the negative correlation between the MG parameter (for example $\Sigma$) and $\sigma_8$ makes the $\sigma_8$ distributions get closer, as can be seen from Fig. \ref{Fig:1MG_constraints} and Fig. \ref{Fig:1MG_sigma8}. Indeed, the fact that SBB+RSD+HST+lowE+DES prefers a negative value of $\Sigma(z=0)-1$ while P18 prefers a positive value slightly alleviates the tension in $\sigma_8$ to about 1.1-$\sigma$. Hence, due to the negative correlation between $\Sigma_0$ and $\sigma_8$ in the model ($\mu=1$, $\Sigma$), the fact that these datasets prefer an opposite sign for $\Sigma_0$ leads to slightly more concordant measurements of $\sigma_8$. However, we point out that then the tension between SBB+RSD+HST+lowE+DES and P18 is due to $\Sigma_0$.
In sum, we find that varying MG parameters may not lead to conclusive results to help alleviate the $\sigma_8$ tension if specific parameterizations are used. For example, setting $\Sigma=1$ or $\eta=1$ increases the original $\sigma_8$ tension. On the other hand, while setting $\mu=1$ causes some shifts in the $\sigma_8$ measurements that diminish the tension, a new tension in $\Sigma_0$ appears. 


\subsection{Model comparison}\label{sec:model-comparison}

The goal of this section is to perform a model comparison between the MG models discussed throughout this work and the $\Lambda$CDM model using a Bayesian model selection approach. This is based on Bayes theorem,
\begin{equation}
P(\boldsymbol{\theta}|\boldsymbol{D},\mathcal{M})=\frac{\mathcal{L}(\boldsymbol{D}|\boldsymbol{\theta},\mathcal{M}) \pi(\boldsymbol{\theta}|\mathcal{M})}{P(\boldsymbol{D}|\mathcal{M})},
\end{equation}
where $\boldsymbol{\theta}$ is the model parameter vector, $\boldsymbol{D}$ represents the data vector, and $\mathcal{M}$ indicates the model.  $P(\boldsymbol{\theta}|\boldsymbol{D},\mathcal{M})$ is the posterior probability and $P(\boldsymbol{D}|\mathcal{M})$ is the Bayesian Evidence.  $\mathcal{L}(\boldsymbol{D}|\boldsymbol{\theta},\mathcal{M})$ denotes the likelihood and $\pi(\boldsymbol{\theta}|\mathcal{M})$ is the prior probability. Among these quantities, it is the Bayesian Evidence that is ultimately useful for model comparison purposes. The Bayes factor is defined as the ratio of model posterior odds divided by the model prior odds, and is described by 
\begin{equation}
\frac{P(\mathcal{M}_i|\boldsymbol{D})}{P(\mathcal{M}_j|\boldsymbol{D})}=           \frac{\pi(\mathcal{M}_i)}{\pi(\mathcal{M}_j)} B_{ij}.
\end{equation}
Here $B_{ij}\equiv \frac{P(\boldsymbol{D}|\mathcal{M}_i)}{P(\boldsymbol{D}|\mathcal{M}_j)}$ is the Bayes factor and the last equation is a consequence of Bayes theorem applied to models. Hence, the Bayes factor is the ratio of Bayesian Evidences which are computed as an integral over the unnormalized posterior, so we can calculate the Bayes Factor as
\begin{equation}
 B_{ij}=\frac{\int d\boldsymbol{\theta}_i \mathcal{L}(\boldsymbol{D}|\boldsymbol{\theta}_i,\mathcal{M}_i) \pi(\boldsymbol{\theta}_i|\mathcal{M}_i)}{\int d\boldsymbol{\theta}_j \mathcal{L}(\boldsymbol{D}|\boldsymbol{\theta}_j,\mathcal{M}_j) \pi(\boldsymbol{\theta}_j|\mathcal{M}_j)},
\end{equation}
where the indices $i$ and $j$ represent two different models to be compared.

We opt to calculate the Bayes factor by using the program MCEvidence from \cite{MCEvidence}, which is a Python package to compute the Bayesian Evidence from MCMC. This package aims to estimate the density of points in the parameter space by calculating the Evidence using the $k$-th nearest neighbour distances in the parameter space through the Mahalanobis distance  \cite{MCEvidence}. MCEvidence works by assuming that the points in the chains are independent and we employ a further marginalization over the calibration parameters used within the Planck plik, commander and SimAll likelihoods (see \cite{Planck2018}) and the typical nuisance parameters from DES such as calibration parameters, galaxy bias, intrinsic alignment and so on (see Table I in \cite{DESMG2018}). Furthermore, we choose to set $k=1$  since the authors of \cite{MCEvidence} argue that this is the optimal choice in terms of accuracy. Moreover, since the Bayesian Evidence analysis depends on the priors chosen, we show in Table \ref{Table:Priors} the priors used for our parameters, which are indeed very wide, as they should be. Here, we point out that when using the parameters $E_{11}$ and $E_{22}$ we assume a time evolution based on $\propto\Omega_{\text{DE}}$ for the MG parameters while $\propto\Omega_{\text{DE}}/\Omega_\Lambda$ if we use $\mu_0$ and $\Sigma_0$, so then we choose wider priors for the first set of parameters. We observe a slightly higher Evidence for $(\mu=1,\eta)$ compared with $(\mu=1,\Sigma)$ in Table \ref{Table:BayesFactor}, which may be associated to prior volume effects. However, we do not observe this trend in some cases where we use DES data since we have to marginalize over a higher number of nuisance parameters, which leads to a greater statistical error on the Evidence \cite{2017PhRvL.119j1301H}. Any preference for ($\mu=1$, $\eta$) over ($\mu=1$, $\Sigma$) or vice-versa in this analysis should be associated to the parameterization and ranges used to fit the data instead of a real physical preference for one parameter or another. However, in general the results from both models were in agreement.

We show in Table \ref{Table:BayesFactor} the results for the computation of the Bayes factor for the models used and the datasets considered in this work. For the interpretation of the model comparison we rely on the Jeffrey's scale using $\ln B_{i0}$ as stated in Table \ref{Table:JeffreyScale}. The index $0$ represents the $\Lambda$CDM model while the index $i$ runs from 1 over the MG models in the order shown in Table \ref{Table:BayesFactor}. We found the correlation length of each chain to be small and performed additional tests with aggressively thinned chains by a factor of 10 and 30. We found that half of the $\ln B_{i0}$ values changed by $<0.4$ and almost all the remaining values changed by $<0.8$ and some few values that include DES changed above the unity. Additionally, for MG models with one MG parameter we compute the $\ln B_{i0}$ values by using the Savage-Dickey density ratio and obtained similar results.

We find that $\Lambda$CDM is the favoured model by the data in the cases when CMBL or DES data are added to P18 or to P18 plus other datasets such as SNe, BAO, and RSD. Hence, $\Lambda$CDM is the favoured model when using lensing data in combination with Planck and other datasets. The only exception is when we use SBB+RSD+HST+lowE where $\Lambda$CDM is still favored even without any lensing, but only marginally. However, some MG parameters are not very well constrained for this dataset combination since high-$\ell$ data from Planck is not used. Also, since some values are $|\ln B_{i 0}|<1$ then this preference is not worth more than a bare mention.

Nevertheless, almost all MG models are marginally favoured over $\Lambda$CDM when using P18 only, and this goes slightly into the moderate range in a couple of cases (see Table  \ref{Table:BayesFactor}). Moreover, if we consider P18 with the addition of other datasets (except gravitational lensing) we find that models with $\eta=1$ or $\mu=1$ are always favoured over $\Lambda$CDM. Among these models, the ($\mu=1$, $\eta$) parameterization stands out, leading to $\ln B_{30} = 2.5$ or $\ln B_{30} = 2.6$ when we use P18+SNe+BAO or P18+SNe+BAO+RSD, respectively. Therefore, there exists a moderate preference for MG models with one MG parameter over $\Lambda$CDM for some datasets, subject to $\Sigma\neq1$. However, adding CMBL or DES to any combination restores the preference to $\Lambda$CDM and GR. 

It is worth mentioning that a similar situation has been reported when using the lensing anomaly parameter $A_\text{L}$ defined in \cite{2008-lensing-anomaly}. This consistency parameter effectively multiplies the CMB lensing potential power spectrum by $C^\Phi_\ell \rightarrow A_\text{L} C^\Phi_\ell$ and then producing a smoothing of the peaks of the CMB temperature power spectrum. While the standard lensing theory requires $A_\text{L}=1$, a preference for a value of $A_\text{L}$ higher than the unity was reported in \cite{Planck2018} when the full TTTEEE+lowE dataset was used. Moreover, this last work also studied the main degeneracy between $\Sigma$ and $A_\text{L}$ while both parameters are varied together (see also \cite{Lensing.Anomaly2} for an extended discussion). It was shown that if $A_\text{L}=1$ the Planck power spectra prefers $\Sigma>1$, while if $A_\text{L}>1$ is allowed then the preference for higher values of $\Sigma$ is reduced. Therefore, the preference for an enhanced lensing signal from MG reported here may be associated with the preference for a smoothing effect of the CMB peaks by Planck, and vice versa. This will be further explored in follow up future analysis. 

\section{Summary and conclusion}

In this analysis we dissected the constraints on MG parameters by varying one parameter at a time while fixing the other at its GR value. This increases the constraining power of the datasets and allows observing the behavior of each parameter individually with respect to its GR value of unity. We also divide datasets into various combinations to see which datasets or subgroups of datasets are consistent with GR and which ones are not. We found the following: 
\begin{itemize}

    \item Constraints on some MG parameters are significantly improved for one-parameter MG models. For example, if we set $\eta=1$ we find an improvement in the constraints of P18 to about 76\% and 32\% in $\mu_0-1$ and $\Sigma_0-1$, respectively. This tightening leads to higher tension with GR than for models with two parameters for some combinations of datasets. 
    
    \item A Bayesian model selection analysis shows that some of the one-parameter MG models are moderately favoured in comparison with $\Lambda$CDM when P18, P18+SNe+BAO(+RSD), or P18+RSD datasets are used, except when $\Sigma$ is fixed to its GR value. When CMBL or DES data is added to any combination, GR and $\Lambda$CDM become the favoured models again. See the previous section for a brief discussion on the relation with the lensing anomaly parameter $A_\text{L}$.
    
   \item For example, the models ($\mu=1$, $\eta$) and ($\mu$, $\eta=1$) exhibit a 3.8-$\sigma$ (Index of Inconsistency IOI= 7.1) and 3.9-$\sigma$ (IOI=7.4) departure from their GR values, respectively, when using the P18+SNe+BAO dataset combination, while the model with ($\mu$, $\eta$) shows a tension of 3.4-$\sigma$ (IOI=5.8). Note that using P18+SNe+BAO+RSD leads to a high tension such as 3.9-$\sigma$ (IOI=7.5) for ($\mu$, $\eta$), but the model is not favoured over $\Lambda$CDM from the Bayesian comparison, which is not strictly the case for ($\mu=1$, $\Sigma$). 
   
   \item P18 data shows a moderate tension with GR and shows a marginal preference for ($\mu=1$, $\eta$) and ($\mu=1$, $\Sigma$) models over $\Lambda$CDM. This tension and Bayesian preference are strengthened for the combinations P18+RSD and P18+SNe+BAO(+RSD). Here again, the tensions and preferences against GR or $\Lambda$CDM go away when adding CMB Lensing and DES data to any of the combinations. 
   
    \item The tension with GR using some datasets seems to be more associated with the MG parameter $\Sigma$ because when setting this to its GR value no tension is shown for the other varied parameter. This is alleviated when adding CMB Lensing and DES data, which are able to constrain this MG parameter. So it remains to be explored with better data in the future if this tension is due to the inability of some datasets to sufficiently constrain this parameter, some systematic effects, or the underlying models.  
    
    \item It is found that some of the MG models with one varying parameter are successful in reducing the $\sigma_8$ tension. However, some models such as $(\mu, \eta=1)$ and $(\mu, \Sigma=1)$ add to the $\sigma_8$ tension (see Fig. \ref{Fig:1MG_sigma8}) so it does not seem to be the case in general that MG models can consistently solve this tension. 
    
\end{itemize}

In sum, it is found that constraining one MG parameter at a time leads to further useful results on constraining MG models and departures from GR. The results here from parameter constraints and Bayesian model comparisons suggest that the tensions with GR need to be explored more closely when more constraining data will become available. It is found that, in general, Planck 2018, SNe, BAO, and RSD have some tension with GR. However, CMB Lensing and DES-Y1 data are found to restore consistency with GR and this seems to be attributed to their ability to constrain the MG parameter $\Sigma$ that enters into the motion of null rays. This dichotomy can be due to systematic effects in some of the datasets, their current constraining power, or some issues with the models. These possibilities should be investigated using more constraining lensing data from the incoming DES-Y3 and, later, from the Vera Rubin Observatory LSST survey, Euclid and the Roman Space Telescope, as well as CMB data from future experiments such as, for example, the Simons Observatory and CMB Stage-4 experiments.


\section*{Acknowledgements}
We thank Michael Kesden for useful comments and Orion Ning for proofreading the manuscript. 
M.I. acknowledges that this material is based upon work supported in part by the Department of Energy, Office of Science, under Award Number DE-SC0019206. C.G.Q. gratefully acknowledges a PhD scholarship from the Mexican National Council for Science and Technology (CONACYT). The authors acknowledge the Texas Advanced Computing Center (TACC) at The University of Texas at Austin for providing HPC resources that have contributed to some of the research results reported within this paper. URL: http://www.tacc.utexas.edu. 

\section*{Data Availability}
The data underlying this article are available in GitHub, at \url{https://github.com/mishakb/ISiTGR}. The datasets were derived from sources in the public domain: Planck 2018 data can be found at \url{http://pla.esac.esa.int/pla/#cosmology} and the rest of the data can be found at \url{https://github.com/cmbant/CosmoMC/tree/master/data}.
%


\bibliographystyle{mnras}
\bibliography{mnras} 




\appendix

\section{\texttt{ISiTGR} Python wrapper}\label{appendix:ISiTGR}

\begin{figure}
{\includegraphics[width=8cm]{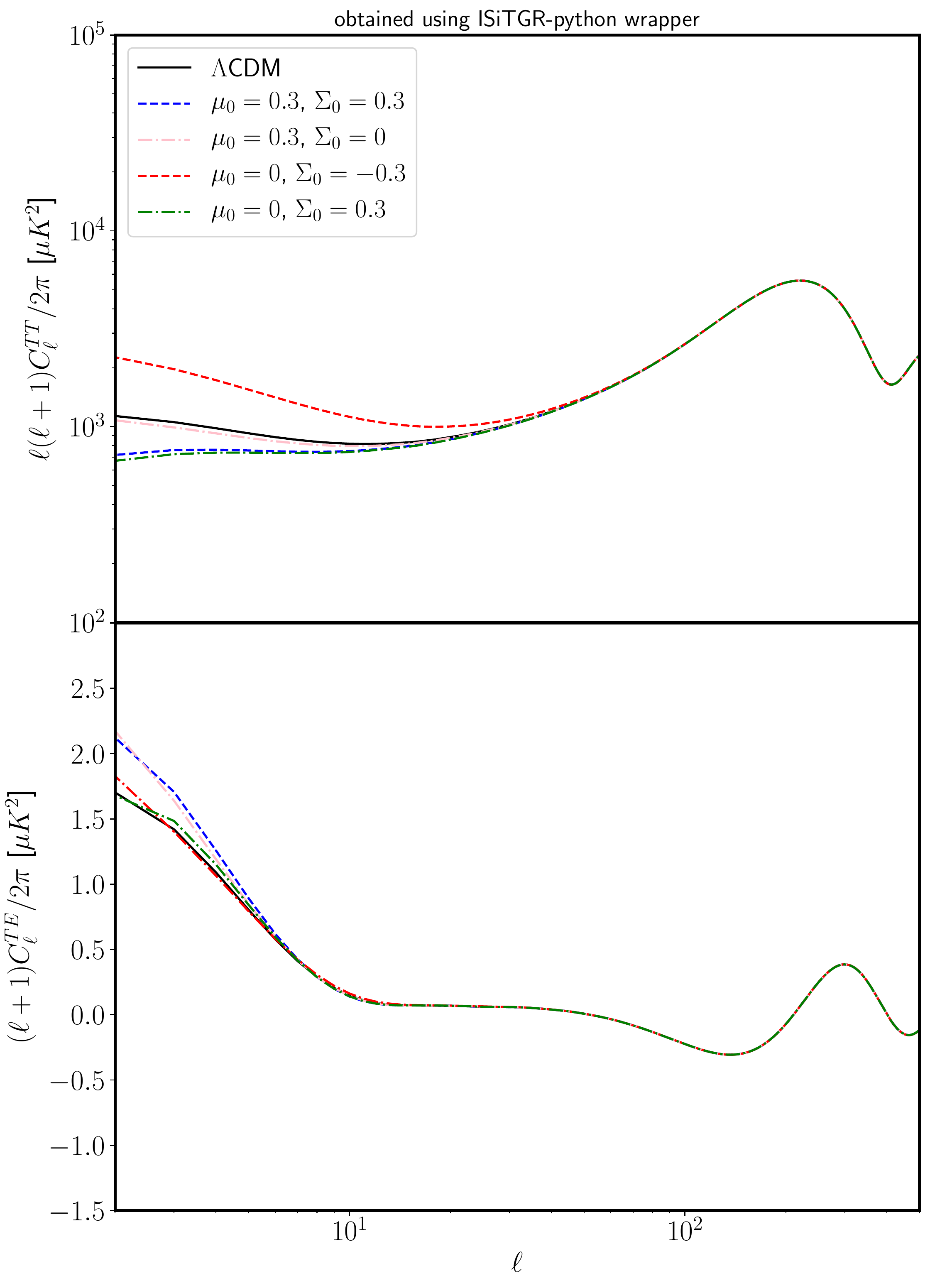}}
\caption{CMB temperature power spectra in the top figure and CMB temperature-polarization power spectra in the bottom. The black curve represents the $\Lambda$CDM model while the other curves show different MG models.}
\label{Fig:CMB-spectra}
\end{figure}
\begin{figure}
{\includegraphics[width=8cm]{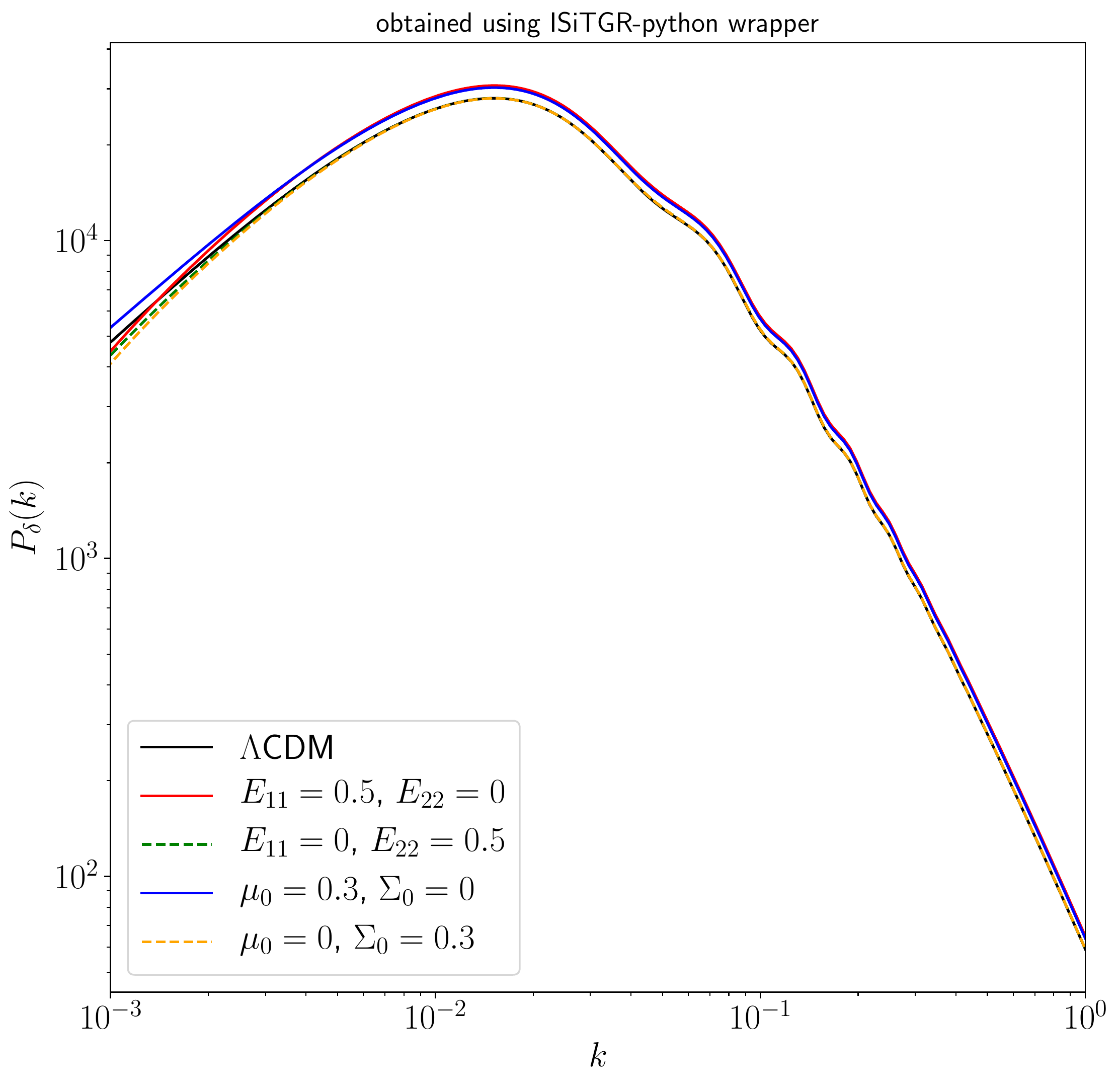}}
\caption{Matter power spectrum for the $\Lambda$CDM model and extended models. The black curve stands for the prediction given by the $\Lambda$CDM model and the other curves represent extensions that include different values of MG parameters. Here, this MG formalism only has an effect at linear scales.}
\label{Fig:MatPow-spectra}
\end{figure}

We present a new Python wrapper for the current version of \texttt{ISiTGR}. The wrapper is intended to make easier and more practical the use of \texttt{ISiTGR} in extending the functionalities of \texttt{CAMB} to include MG parameters. Therefore, the user is allowed to obtain transfer functions as well as matter, lensing, CMB power spectra and other cosmological calculations. We should remind the reader that the Python wrapper provides cosmological theoretical calculations usually provided by \texttt{CAMB}. However, if the user plans to perform MCMC sampling analysis, we recommend using the full Fortran-based version of \texttt{ISiTGR}, which is a patch to both \texttt{CAMB} and \texttt{CosmoMC}. 
More information about the \texttt{ISiTGR} Python wrapper together with examples in a Jupyter notebook can be found at \url{https://isitgr.readthedocs.io/en/latest/}. Instructions about how to modify the source code to include your own parameterizations and implement them into the Python wrapper can be found at ~\url{https://github.com/mishakb/ISiTGR} where there is more information on the entire \texttt{ISiTGR} system  \cite{ISITGR,ISiTGR-CGQ}.
A special effort was made to make \texttt{ISiTGR} user-friendly with simple installation instruction. The Python wrapper was added to further facilitate its use.  

In order to show some of the capabilities of the new \texttt{ISiTGR} Python wrapper, we plot in Fig. \ref{Fig:CMB-spectra} the CMB temperature and temperature-polarization power spectra for $\Lambda$CDM and various MG models as obtained from the wrapper. As we can see, for the temperature power spectra, the deviations from GR due to the MG parameters are significant at $\ell<100$. On the other hand, the modifications in the temperature-polarization power spectra occur at $\ell<10$, which is a region dominated by cosmic variance. Furthermore, we plot the matter power spectrum in Fig. \ref{Fig:MatPow-spectra} for models with only one MG parameter. As we can see, models with $\mu \neq 1$ affect the matter power spectrum at all scales, while models with $\mu=1$ but with some other MG parameter different from its GR value may have an effect only at very large scales (small $k$). 

As a summary, the \texttt{ISiTGR} Python wrapper provides extensions to $\Lambda$CDM based on four different MG phenomenological parameterizations, which include time-dependent as well as scale-dependent parameterizations. It also allows the user to make use of both functional form and binning methods to calculate the MG parameters. Additionally, \texttt{ISiTGR} works with spatially flat or curved universes.


\bsp	
\label{lastpage}
\end{document}